\documentclass[aps,prd,twocolumn,superscriptaddress]{revtex4-2}

\usepackage{siunitx}
\usepackage{orcidlink}
\usepackage{enumitem}
\usepackage{diagbox}
\usepackage{lineno}
\usepackage{natbib}
\usepackage{xspace}

\usepackage{etoolbox} 
\newcommand{\AckText}{}
\newcommand{\AddAck}[1]{\appto{\AckText}{#1\space}}

\newcommand{\pastro}{\ensuremath{p_{\mathrm{astro}}}\xspace}

\begin{document}


\title{GstLAL O4 Online Results}

\author{Shomik Adhicary \orcidlink{0009-0004-2101-5428}}
\email{shomik.adhicary@ligo.org}
\affiliation{Department of Physics, The Pennsylvania State University, University Park, PA 16802, USA}
\affiliation{Institute for Gravitation and the Cosmos, The Pennsylvania State University, University Park, PA 16802, USA}

\author{Pratyusava Baral \orcidlink{0000-0001-6308-211X}}
\affiliation{Leonard E.\ Parker Center for Gravitation, Cosmology, and Astrophysics, University of Wisconsin-Milwaukee, Milwaukee, WI 53201, USA}

\author{Amanda Baylor \orcidlink{0000-0003-0918-0864}}
\affiliation{Leonard E.\ Parker Center for Gravitation, Cosmology, and Astrophysics, University of Wisconsin-Milwaukee, Milwaukee, WI 53201, USA}

\author{Becca Ewing}
\affiliation{Department of Physics, The Pennsylvania State University, University Park, PA 16802, USA}
\affiliation{Institute for Gravitation and the Cosmos, The Pennsylvania State University, University Park, PA 16802, USA}

\author{Yun-Jing Huang \orcidlink{0000-0002-2952-8429}}
\affiliation{Department of Physics, The Pennsylvania State University, University Park, PA 16802, USA}
\affiliation{Institute for Gravitation and the Cosmos, The Pennsylvania State University, University Park, PA 16802, USA}

\author{Rachael Huxford}
\affiliation{Minnesota Supercomputing Institute, University of Minnesota, Minneapolis, MN 55455, USA}

\author{Prathamesh Joshi \orcidlink{0000-0002-4148-4932}}
\affiliation{Department of Physics, The Pennsylvania State University, University Park, PA 16802, USA}
\affiliation{Institute for Gravitation and the Cosmos, The Pennsylvania State University, University Park, PA 16802, USA}
\affiliation{School of Physics, Georgia Institute of Technology, Atlanta, GA 30332, USA}

\author{James Kennington \orcidlink{0000-0002-6899-3833}}
\affiliation{Department of Physics, The Pennsylvania State University, University Park, PA 16802, USA}
\affiliation{Institute for Gravitation and the Cosmos, The Pennsylvania State University, University Park, PA 16802, USA}

\author{Ryan Magee \orcidlink{0000-0001-9769-531X}}
\affiliation{LIGO Laboratory, California Institute of Technology, Pasadena, CA 91125, USA}

\author{Cody Messick \orcidlink{0000-0002-8230-3309}}
\affiliation{Leonard E.\ Parker Center for Gravitation, Cosmology, and Astrophysics, University of Wisconsin-Milwaukee, Milwaukee, WI 53201, USA}

\author{Wanting Niu \orcidlink{0000-0003-1470-532X}}
\affiliation{Department of Physics, The Pennsylvania State University, University Park, PA 16802, USA}
\affiliation{Institute for Gravitation and the Cosmos, The Pennsylvania State University, University Park, PA 16802, USA}

\author{Cort Posnansky \orcidlink{0009-0009-7137-9795}}
\affiliation{Department of Physics, The Pennsylvania State University, University Park, PA 16802, USA}
\affiliation{Institute for Gravitation and the Cosmos, The Pennsylvania State University, University Park, PA 16802, USA}

\author{Surabhi Sachdev \orcidlink{0000-0002-0525-2317}}
\affiliation{School of Physics, Georgia Institute of Technology, Atlanta, GA 30332, USA}

\author{Shio Sakon \orcidlink{0000-0002-5861-3024}}
\affiliation{Department of Physics, The Pennsylvania State University, University Park, PA 16802, USA}
\affiliation{Institute for Gravitation and the Cosmos, The Pennsylvania State University, University Park, PA 16802, USA}
\AddAck{S.S. was supported by College Women's Association Japan Graduate Scholarship for Women to Study Abroad}

\author{Urja Shah \orcidlink{0000-0001-8249-7425}}
\affiliation{School of Physics, Georgia Institute of Technology, Atlanta, GA 30332, USA}

\author{Divya Singh \orcidlink{0000-0001-9675-4584}}
\affiliation{Department of Physics, University of California, Berkeley, CA 94720, USA}
\affiliation{Institute for Nuclear Theory, University of Washington, Seattle, WA 98195, USA}
\AddAck{DS acknowledges support from NSF Grant PHY-2020275(Network for Neutrinos, Nuclear Astrophysics, and Symmetries (N3AS)).}

\author{Leo Tsukada  \orcidlink{0000-0003-0596-5648}}
\affiliation{Department of Physics and Astronomy, University of Nevada, Las Vegas, 4505 South Maryland Parkway, Las Vegas, NV 89154, USA}
\affiliation{Nevada Center for Astrophysics, University of Nevada, Las Vegas, NV 89154, USA}
\AddAck{LT acknowledges NASA 80NSSC23M0104 and the Nevada Center for Astrophysics for support.}

\author{Zach Yarbrough \orcidlink{0000-0002-9825-1136}}
\affiliation{Department of Physics and Astronomy, Louisiana State University, Baton Rouge, LA 70803, USA}

\author{Noah Zhang \orcidlink{0009-0003-3361-5538}}
\affiliation{School of Physics, Georgia Institute of Technology, Atlanta, GA 30332, USA}

\author{Kipp Cannon \orcidlink{0000-0003-4068-6572}}
\affiliation{RESCEU, The University of Tokyo, Tokyo, 113-0033, Japan}

\author{Sarah Caudill}
\affiliation{Department of Physics, University of Massachusetts, Dartmouth, MA 02747, USA}
\affiliation{Center for Scientific Computing and Data Science Research, University of Massachusetts, Dartmouth, MA 02747, USA}

\author{Bryce Cousins \orcidlink{0000-0002-7026-1340}}
\affiliation{Department of Physics, University of Illinois, Urbana, IL 61801 USA}
\affiliation{Department of Physics, The Pennsylvania State University, University Park, PA 16802, USA}
\affiliation{Institute for Gravitation and the Cosmos, The Pennsylvania State University, University Park, PA 16802, USA}
\AddAck{B.C. acknowledges support from the NSF Graduate Research Fellowship Program under Grant No. DGE 21-46756.}

\author{Jolien D. E. Creighton \orcidlink{0000-0003-3600-2406}}
\affiliation{Leonard E.\ Parker Center for Gravitation, Cosmology, and Astrophysics, University of Wisconsin-Milwaukee, Milwaukee, WI 53201, USA}

\author{Heather Fong}
\affiliation{Department of Physics and Astronomy, University of British Columbia, Vancouver, BC, V6T 1Z4, Canada}
\affiliation{RESCEU, The University of Tokyo, Tokyo, 113-0033, Japan}
\affiliation{Graduate School of Science, The University of Tokyo, Tokyo 113-0033, Japan}

\author{Richard N. George \orcidlink{0000-0002-7797-7683}}
\affiliation{Center for Gravitational Physics, University of Texas at Austin, Austin, TX 78712, USA}

\author{Olivia Godwin \orcidlink{0000-0002-7489-4751}}
\affiliation{LIGO Laboratory, California Institute of Technology, MS 100-36, Pasadena, California 91125, USA}
\affiliation{Department of Physics, The Pennsylvania State University, University Park, PA 16802, USA}
\affiliation{Institute for Gravitation and the Cosmos, The Pennsylvania State University, University Park, PA 16802, USA}

\author{Reiko Harada}
\affiliation{RESCEU, The University of Tokyo, Tokyo, 113-0033, Japan}
\affiliation{Graduate School of Science, The University of Tokyo, Tokyo 113-0033, Japan}

\author{Soichiro Kuwahara}
\affiliation{RESCEU, The University of Tokyo, Tokyo, 113-0033, Japan}
\affiliation{Graduate School of Science, The University of Tokyo, Tokyo 113-0033, Japan}

\author{Alvin K. Y. Li \orcidlink{0000-0001-6728-6523}}
\affiliation{RESCEU, The University of Tokyo, Tokyo, 113-0033, Japan}
\affiliation{Graduate School of Science, The University of Tokyo, Tokyo 113-0033, Japan}

\author{Duncan Meacher \orcidlink{0000-0001-5882-0368}}
\affiliation{Leonard E.\ Parker Center for Gravitation, Cosmology, and Astrophysics, University of Wisconsin-Milwaukee, Milwaukee, WI 53201, USA}

\author{Soichiro Morisaki \orcidlink{0000-0002-8445-6747}}
\affiliation{Institute for Cosmic Ray Research, The University of Tokyo, 5-1-5 Kashiwanoha, Kashiwa, Chiba 277-8582, Japan}

\author{Debnandini Mukherjee  \orcidlink{0000-0001-7335-9418}}
\affiliation{School of Physics and Astronomy, University of Birmingham, Edgbaston, Birmingham, B15 2TT, United Kingdom}
\affiliation{Institute for Gravitational Wave Astronomy, University of Birmingham, Edgbaston, Birmingham, B15 2TT, United Kingdom}

\author{Alexander Pace \orcidlink{0009-0003-4044-0334}}
\affiliation{Department of Physics, The Pennsylvania State University, University Park, PA 16802, USA}
\affiliation{Institute for Gravitation and the Cosmos, The Pennsylvania State University, University Park, PA 16802, USA}

\author{Anarya Ray \orcidlink{0000-0002-7322-4748}}
\affiliation{Leonard E.\ Parker Center for Gravitation, Cosmology, and Astrophysics, University of Wisconsin-Milwaukee, Milwaukee, WI 53201, USA}
\affiliation{Center for Interdisciplinary Exploration and Research in Astrophysics, Northwestern University, IL 60201, USA}

\author{Stefano Schmidt \orcidlink{0000-0002-8206-8089}}
\affiliation{Nikhef, Science Park 105, 1098 XG, Amsterdam, The Netherlands}
\affiliation{Institute for Gravitational and Subatomic Physics (GRASP), Utrecht University, Princetonplein 1, 3584 CC Utrecht, The Netherlands}

\author{Ron Tapia}
\affiliation{Department of Physics, The Pennsylvania State University, University Park, PA 16802, USA}
\affiliation{Institute for Computational and Data Sciences, The Pennsylvania State University, University Park, PA 16802, USA}

\author{Koh Ueno \orcidlink{0000-0003-3227-6055}}
\affiliation{RESCEU, The University of Tokyo, Tokyo, 113-0033, Japan}

\author{Aaron Viets \orcidlink{0000-0002-4241-1428}}
\affiliation{Concordia University Wisconsin, Mequon, WI 53097, USA}

\author{Leslie Wade}
\affiliation{Department of Physics, Hayes Hall, Kenyon College, Gambier, Ohio 43022, USA}

\author{Madeline Wade \orcidlink{0000-0002-5703-4469}}
\affiliation{Department of Physics, Hayes Hall, Kenyon College, Gambier, Ohio 43022, USA}

\author{Graham Woan \orcidlink{0000-0003-0381-0394}}
\affiliation{SUPA, University of Glasgow, Glasgow G12 8QQ, United Kingdom}

\author{Chad Hanna}
\affiliation{Department of Physics, The Pennsylvania State University, University Park, PA 16802, USA}
\affiliation{Institute for Gravitation and the Cosmos, The Pennsylvania State University, University Park, PA 16802, USA}
\affiliation{Department of Astronomy and Astrophysics, The Pennsylvania State University, University Park, PA 16802, USA}
\affiliation{Institute for Computational and Data Sciences, The Pennsylvania State University, University Park, PA 16802, USA}
\AddAck{
CH Acknowledges generous support from the Eberly College of Science, the
Department of Physics, the Institute for Gravitation and the Cosmos, the
Institute for Computational and Data Sciences, and the Freed Early Career Professorship.}


\date{\today}

\begin{abstract}
Gravitational-wave observations of merging binary neutron stars and black holes are now routinely made by detectors in the Advanced LIGO--Virgo--KAGRA network.
Neutron star binary systems may also produce detectable electromagnetic and particle emission over times scales ranging from seconds to years.
Real-time gravitational-wave searches play a central role in enabling time-critical electromagnetic and/or neutrino follow-up observations.
During the fourth observing run (O4) of the Advanced LIGO--Virgo--KAGRA network, multiple real-time searches operated continuously to identify candidate gravitational-wave events and publicly disseminate information about these discoveries.
Here, the performance and results of the GstLAL real-time analysis are reported.
The analysis is designed to identify candidates with low latency, high detection efficiency, and sustained operational uptime over long observing periods.
Across O4, it produced initial candidate uploads with a median latency of \qty{15.8}{\second} while maintaining an effective uptime of 98\% during the first two parts of the observing run.
During the run, the analysis contributed to 250 candidates classified as astrophysically plausible, provided the first upload for 222 of these, and was the sole contributor for 75.
Among Gravitational-Wave Transient Catalog events with a false-alarm rate below one per year, 88\% were identified as significant in low latency and promoted for expert vetting and public dissemination.
The low-latency astrophysical classifications agreed with the final catalog classifications for 93\% of the events considered.
\end{abstract}


\maketitle


\section{Introduction}

Modern astronomy benefits from observations across multiple complementary messengers, including electromagnetic radiation, neutrinos, and gravitational-waves (GWs) \cite{Meszaros:2019mma}.
GWs provide direct information about compact astrophysical systems consisting of black holes and neutron stars \cite{LIGOScientific:2016aoc, LIGOScientific:2024elc, LIGOScientific:2017vwq}.
GW observations can trigger rapid electromagnetic and neutrino follow-up observations, forming the basis of multi-messenger astronomy \cite{LIGOScientific:2017ync, ANTARES:2017bia}.
Realizing multi-messenger astronomy requires candidate GW events to be identified and distributed within seconds to minutes of their occurrence \cite{Magee:2021xdx, gwcelery_docs, Cabero:2020eik, LIGOScientific:2019gag}.

Low-latency GW detection pipelines are designed to identify candidate events in near real time and to provide basic information needed for rapid follow-up \cite{Messick:2016aqy, DalCanton:2020vpm, Klimenko:2015ypf, Allene:2025saz, Chu:2020pjv}.
The information \cite{LVK_EMFollow_UserGuide_Content} includes the estimated time at which the GW signal passed through the detectors, an approximate region of the sky from which it originated \cite{Singer:2015ema}, and a coarse classification of the source \cite{Farr:2013yna, Kapadia:2019uut}.
The source classification indicates whether the signal is consistent with a merger of two black holes, two neutron stars, or a black hole--neutron star system, based on broad features of the inferred component masses and spins.
Low-latency GW detection pipelines run continuously during observing periods and operate under constraints imposed by finite computing resources and data that may still contain instrumental artifacts \cite{Davis:2022dnd}.
As observing runs have increased in duration and the rate of detected signals has grown \cite{LIGOScientific:2025hdt}, maintaining stable low-latency analyses over long timescales has become an increasingly common requirement for real-time GW searches.

GstLAL is a matched-filtering search pipeline used for the detection of GWs from compact binary coalescences (CBCs) \cite{Cannon:2011rj, Messick:2016aqy, Hanna:2019ezx, Sachdev:2019vvd, Cannon:2020qnf, Sakon:2022ibh, Tsukada:2023edh, Ewing:2023qqe, Ray:2023nhx, Joshi:2025nty, Joshi:2025zdu}.
Matched filtering identifies signals by correlating detector data against a bank of waveform templates that model expected GW signals \cite{Owen:1998dk, Allen:2005fk}.
Candidate triggers from individual detectors are combined to form events using multi-detector coincidence, which requires consistency in time and signal parameters across the detector network \cite{Cannon:2015gha, Messick:2016aqy, Hanna:2019ezx}.
Coincident events are assigned a ranking statistic that summarizes their likelihood of being astrophysical relative to the measured background \cite{Sachdev:2019vvd, Tsukada:2023edh}.
Additional astrophysical classification is performed to estimate the probability that a candidate belongs to broad source categories \cite{Ray:2023nhx}.

The GstLAL pipeline operates in both online (low-latency) and offline (longer-latency, archival) configurations \cite{Ewing:2023qqe, Joshi:2025zdu}.
The online analysis prioritizes rapid event identification and reporting, while the offline analysis reprocesses data with relaxed latency constraints and is used for the construction of the Gravitational-Wave Transient Catalog (GWTC) \cite{LIGOScientific:2018mvr, LIGOScientific:2020ibl, LIGOScientific:2021usb, KAGRA:2021vkt, LIGOScientific:2025slb}.
The GWTC compiles the set of all astrophysically plausible GW events identified after detailed offline analysis and review \cite{LIGOScientific:2025hdt}.
The GWTC is updated after each observing run, with successive releases incorporating newly observed events and refined analyses of previously reported signals \cite{LIGOScientific:2025slb}.
GWTC production involves reanalyzing the recorded detector data without real-time constraints, refining estimates of event significance and source properties \cite{LIGOScientific:2025yae}.

During the fourth observing run (O4) of the LIGO Scientific \cite{LIGOScientific:2014pky}, Virgo \cite{VIRGO:2014yos} and KAGRA \cite{KAGRA:2020tym} Collaboration (LVK), the GstLAL online analysis produced low-latency event uploads to the GRAvitational-wave Candidate Event Database (GraceDB) \cite{GraceDB2014} and reported preliminary sky localization and source classification for public alert candidates.
The online analysis results were subsequently reused in downstream workflows associated with the production of the GWTC-4.0 \cite{Joshi:2025zdu, LIGOScientific:2025yae}.
The reuse of online analysis results during O4 represents a change relative to earlier observing runs, in which online and offline analyses were more clearly separated \cite{KAGRA:2021vkt}.
O4 therefore provides an opportunity to examine the behavior of the online analysis operating continuously over an extended period while contributing to both real-time alerts and GWTC production.

The purpose of this work is to evaluate the performance of the GstLAL online analysis during O4, in the context of its intended role within low-latency detection and catalog production.
In O4, the online analysis was configured to operate as a real-time search with sensitivity approaching that of offline analyses \cite{Ewing:2023qqe, Joshi:2025zdu}, while maintaining a reasonable false-alarm rate (FAR) \cite{Cannon:2015gha}.
A central objective is to assess whether the resulting online significance estimates and source classifications remained broadly consistent with those obtained from offline reprocessing used in GWTC-4.0 construction \cite{Joshi:2025nty, LIGOScientific:2025yae}.
This work examines differences between online and offline event identification, ranking, and characterization, as well as the extent to which the online analysis contributed to the formation of superevents \cite{LVK_EMFollow_UserGuide_Superevents} and public alerts.
Performance is evaluated using quantitative metrics including alert latency, analysis uptime, event production and upload rates, FAR and source-classification consistency, and the frequency with which GstLAL provided the first or preferred event \cite{LVK_EMFollow_UserGuide_PreferredEvent} within a superevent.
Edge cases encountered during O4, including events missed online but recovered offline and superevents that were later retracted, are also examined to clarify the limitations of the online configuration and identify areas for future improvement.

The paper is organized to present both the technical performance of the GstLAL online analysis and the operational practices supporting sustained use during O4.
Section~\ref{sec:configuration} describes the configuration of the GstLAL online search and changes relative to earlier observing runs.
Section~\ref{sec:performance} presents quantitative performance metrics that characterize latency, stability, event production, and the accuracy of low-latency quantities.
Section~\ref{sec:experience} focuses on operational experience during O4, including challenges encountered in long-duration real-time operation and the role of targeted automation tools.
Section~\ref{sec:lessons} summarizes the lessons learned during O4 and discusses implications for improving scalability, robustness, and operational sustainability in preparation for the next observing run.
\section{Overview of the GstLAL Online Analysis in O4}
\label{sec:configuration}

Section~\ref{sec:configuration} summarizes the configuration of the GstLAL online analysis during O4, emphasizing only those aspects that are necessary for interpreting the performance metrics presented in Section~\ref{sec:performance}, with algorithmic details deferred to prior work.
A detailed explanation of the methods of the GstLAL online analysis can be found in \cite{Ewing:2023qqe}.

CBC searches aim to identify GW signals from merging compact binaries by filtering detector strain data against large banks of modeled waveforms and assessing the statistical significance of coincident triggers across the detector network. 
CBC searches operate in the presence of non-Gaussian and non-stationary detector noise, requiring robust background estimation and ranking methods to distinguish astrophysical signals from instrumental artifacts while covering a high-dimensional parameter space of masses, spins, orbital configurations and sky positions.

In addition to identifying statistically significant candidates, CBC searches provide basic characterization of each event, including estimates of the event time, sky localization, and a coarse source classification based on inferred component masses and spins.
In the low-latency setting, these outputs are produced under strict computational and timing constraints to support rapid follow-up observations.

\subsection{Pipeline Summary}
\label{sec:pipeline_summary}

The online analysis relies on a precomputed template bank \cite{Sakon:2022ibh, Hanna:2022zpk}, consisting of a discrete set of modeled CBC waveforms that span the stellar-mass CBC parameter space \cite{LIGOScientific:2025yae}.
The templates represent compact binaries on quasi-circular orbits with spin components aligned or anti-aligned with the orbital angular momentum, and include only the dominant quadrupolar $(2,2)$ mode.
The template bank is interleaved into two complementary halves, referred to as checkerboards, which each cover the same parameter space and operate independently at separate computing sites to improve fault tolerance and ensure robustness against hardware or site-level failures.
Each checkerboard is further subdivided into many independent template bins that process the same strain data from the detector network in parallel, with each bin handled by a separate analysis job \cite{Cannon:2011rj}.
As a result, the analysis can produce multiple triggers associated with a single astrophysical signal.
A trigger is a threshold-crossing feature identified in the matched-filter output that may be associated with a CBC signal \cite{Messick:2016aqy}.
Triggers are aggregated and processed to form candidate events, which may then be uploaded to GraceDB \cite{Joshi:2025nty}.

Within the parallel, streaming workflow, the time required to produce and upload candidate events is set by both the incremental processing of incoming data and the subsequent aggregation of triggers across jobs.
As calibrated strain data are recorded by the detectors, they are transferred to the computing sites where the GstLAL online analysis is deployed \cite{LIGOScientific:2025snk}.
The online analysis processes the incoming strain data incrementally in short segments, allowing data conditioning, matched filtering, and candidate identification to proceed continuously \cite{Messick:2016aqy, Ewing:2023qqe}.
Candidate identification \cite{Cannon:2015gha, Messick:2016aqy, Sachdev:2019vvd, Tsukada:2023edh} therefore occurs shortly after the coalescence time, once sufficient data have been accumulated to form a statistically meaningful matched-filter signal-to-noise ratio (SNR).
The streaming architecture introduces latency that reflects data transfer, data conditioning, filtering, formation of coincidence across detectors, significance calculation, and upload to GraceDB.

Candidate significance in GstLAL is quantified using a likelihood-ratio (LR) ranking statistic informed by trigger properties, detector properties, and prior assumptions about the CBC population \cite{Cannon:2015gha, Messick:2016aqy, Tsukada:2023edh}.
The background noise distribution is estimated by accumulating ranking-statistic information from single-detector triggers observed during coincident time, while excluding times around manually-vetted GW events to reduce astrophysical signal contamination.
Background state is cumulative in time and is snapshotted to disk every four hours to support persistent operation and recovery.
Within a specified time window, the candidate with the highest LR across all template bins is selected for FAR estimation \cite{Joshi:2025nty}.
For FAR estimation, bin-level background histograms are marginalized by summing counts across all template bins in a continuous loop that requires hours per iteration and accumulates as observing time grows.
A trials factor of 2 is applied in FAR estimation to reflect the two checkerboarded halves of the analysis.

Candidates selected from aggregated triggers with FARs below a nominal threshold are uploaded to GraceDB.
For each uploaded candidate, the analysis also computes a probability of astrophysical origin, $p_{\mathrm{astro}}$, along with source-class probabilities \cite{Ray:2023nhx}.
The $p_{\mathrm{astro}}$ calculation combines the candidate’s ranking statistic with CBC population models.
$p_{\mathrm{astro}}$ depends on the available background and population assumptions at the time of evaluation.

These stages directly determine the latency, FAR, and $p_{\mathrm{astro}}$-related metrics reported in Sec.~\ref{sec:performance}.

During O4, the GstLAL online framework also supported early-warning, subsolar-mass, and exploratory low-latency analyses running in parallel with the stellar-mass search.
However, this work reports only the performance and results of the stellar-mass analysis.

%
%

\subsection{Online Infrastructure}

The analysis relies on a centralized monitoring framework that collects health information from individual analysis jobs and uses it to track overall system status and detect failures in real time.
The framework is built from several complementary components, each responsible for a different aspect of monitoring analysis health.

At the top level, operational alerting is handled by Icinga \cite{icinga}, a service-monitoring system that continuously evaluates whether predefined operating conditions are being met.
For each active GstLAL online analysis, Icinga runs a set of service checks that test whether critical processes are running, responsive, and producing expected outputs.
Checks are organized by analysis and individual jobs, allowing responders to quickly localize problems, particularly during periods of high activity.

Time-dependent performance and liveness information from individual jobs is collected using InfluxDB \cite{influxdata}, a time-series database designed to store and query timestamped metrics.
Each job periodically reports basic internal state information to InfluxDB, such as heartbeat signals and limited resource usage indicators.
Icinga queries these metrics to confirm that jobs remain active and advancing in time, with thresholds on metric recency to identify stalled or terminated jobs.

In addition to metric-based monitoring, each analysis job exposes a lightweight health interface via a Hypertext Transfer Protocol (HTTP) endpoint.
These endpoints provide a direct indication of job responsiveness and internal consistency.
Icinga performs regular HTTP checks against these interfaces, allowing it to detect failures that may not immediately appear in time-series metrics, such as partial hangs or internal logic errors.
Using both InfluxDB-based and HTTP-based checks provides redundancy across different failure modes.

Monitoring signals generated by Icinga are consumed by automated systems and human responders.
Corrective actions, such as restarting or resubmitting failed jobs, are handled by separate control and scheduling infrastructure based on HTCondor \cite{htcondor}.
In the online analysis workflow, analysis jobs are organized as directed acyclic graphs (DAGs), in which nodes represent individual processing jobs and edges encode execution dependencies between them.
The separation allows monitoring to focus on detection and diagnosis, while remediation is delegated to systems designed for large-scale job control.

\begin{figure}
\includegraphics[width=\columnwidth]{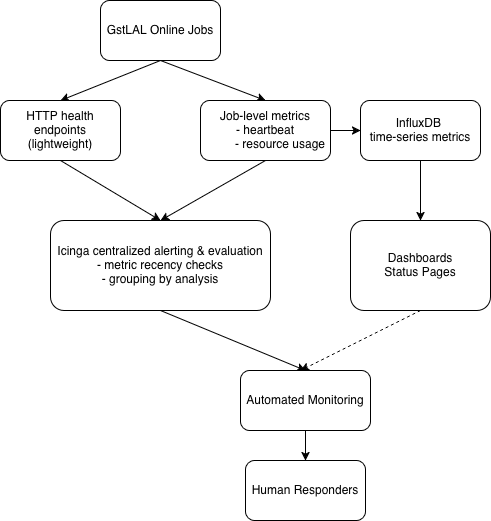}
\caption{
    Overview of the monitoring and alerting infrastructure used by the GstLAL online analysis during O4.
    Job-level metrics, HTTP health endpoints, and analysis-level service checks are evaluated by Icinga services to provide centralized monitoring and real-time alerting to automated systems and human responders.
}
\label{fig:monitoring}
\end{figure}

An overview of the monitoring and alerting infrastructure used by the GstLAL online analysis is shown in Fig.~\ref{fig:monitoring}.


\subsection{Run Timeline}

The O4 observing run was divided into three primary parts, O4a, O4b, and O4c, each associated with changes in detector availability, analysis configuration, and computing deployment \cite{LIGOScientific:2025hdt}.
O4a spanned from May 4, 2023 to January 16, 2024, while O4b covered the period from April 10, 2024 to January 28, 2025.
O4c followed shortly after, beginning on January 28, 2025 and continuing through November 18, 2025, with a gap separating the two observing parts within O4c.

During O4a, two stellar-mass analyses were operated in parallel, named \emph{Edward} and \emph{Jacob}.
These analyses corresponded to the two GstLAL checkerboards, as described in Sec.~\ref{sec:pipeline_summary} and processed data from the Hanford–Livingston (HL) detector network.
Edward was hosted at the CIT computing site, while Jacob ran at ICDS, providing redundancy and operational resilience.

The O4b configuration expanded both detector coverage and computing footprint.
Edward and Jacob continued to operate, while two additional analyses, \emph{Rick} and \emph{Bob}, were introduced.
Rick was deployed at ICDS and Bob at the NEMO cluster, with both analyses processing data from the Hanford–Livingston–Virgo (HLV) network.
Midway through O4b, Edward and Jacob were retired in favor of Rick and Bob, reflecting improved sensitivity of the newer configurations.

O4c began only a few hours after the conclusion of O4b.
Background statistics accumulated by Rick and Bob during O4b were reused to initialize the O4c analyses, enabling continuity in sensitivity while keeping data products from different observing segments logically separated.
This approach reduced the time required to achieve stable background estimation at the start of O4c.

Throughout O4, new features, tuning updates, and bug fixes were incorporated incrementally.
Such changes were validated using staging and testing analyses that ran in parallel with production pipelines.
These parallel analyses processed both historical data and newly acquired data, allowing improvements to be deployed without disrupting ongoing low-latency operations.

\section{Quantitative Performance Metrics}
\label{sec:performance}

Section~\ref{sec:performance} presents quantitative performance metrics that characterize the operation of the GstLAL online analysis during O4. 
The discussion is organized around four complementary aspects: uptime and duty cycle of the online analysis, latency of candidate uploads, event production and GstLAL’s contribution to ADVOK superevents, and the accuracy of low-latency quantities such as \pastro, and FAR relative to offline analyses. 
Together, the metrics translate the configuration choices described in Sec.~\ref{sec:configuration} into measurable performance, enabling an assessment of how reliably the analysis delivered rapid, astrophysically meaningful alerts throughout the run. 
The section concludes with a focused examination of superevents that did not cross the significant alert threshold \cite{emfollow_far} online but were recovered with higher significance offline. 
Together with the earlier metrics, that examination clarifies the strengths and limitations of the O4 online configuration.

\subsection{Uptime and Duty Cycle}

\begin{figure}
\includegraphics[width=\columnwidth]{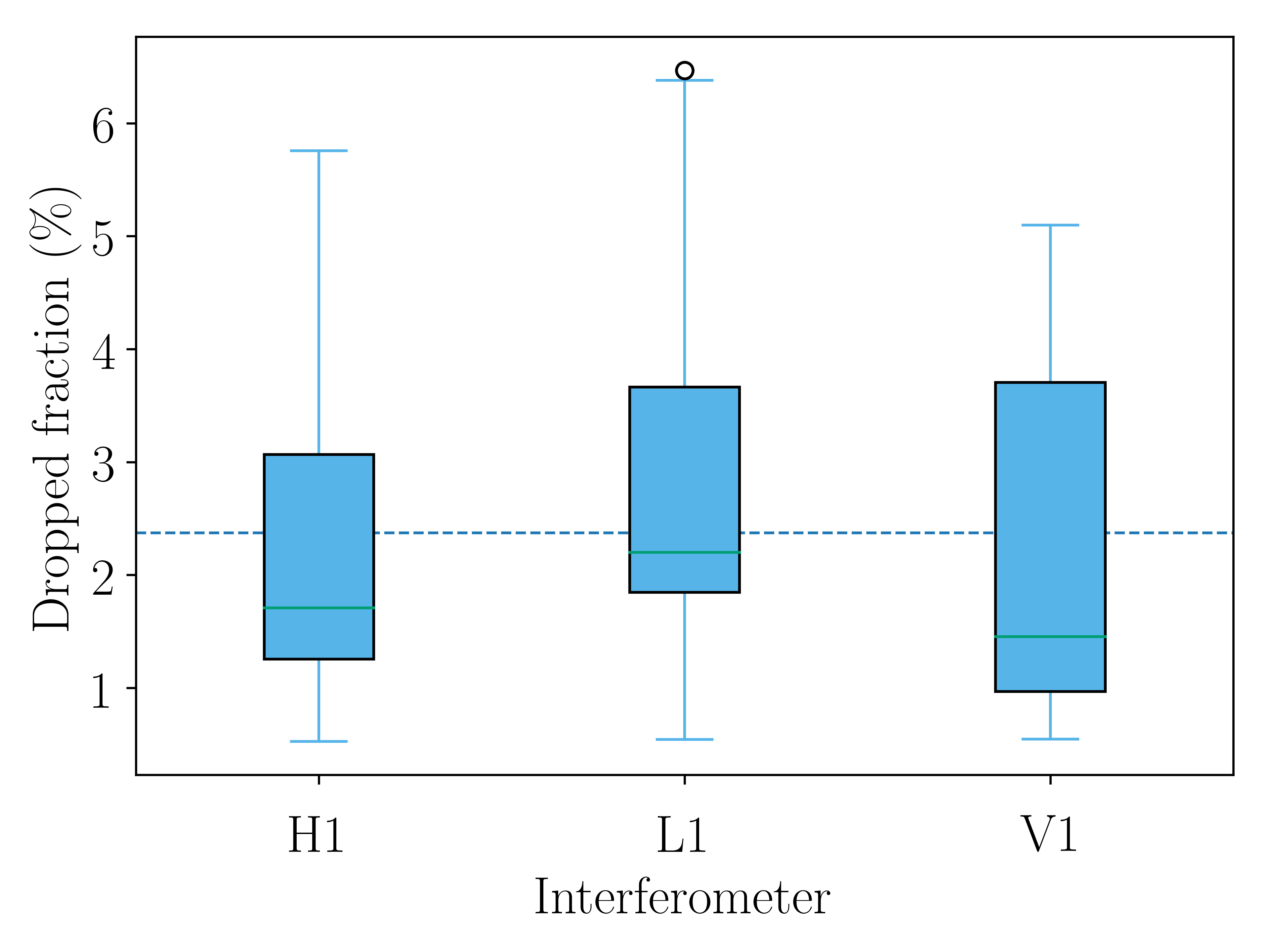}
\caption{
    Distribution of the percentage of data dropped by independent HTCondor jobs for each IFO analyzed by the GstLAL online analysis during O4.
    Each point represents a job within one of the two checkerboards.
    The box shows the interquartile range with the median indicated by the horizontal line.
    Whiskers extend to 1.5 × IQR, and outliers are shown as points.
    The low dropped fraction (typically below \qty{7}{\percent}) indicates an overall analysis uptime exceeding \qty{93}{\percent}.
    The dashed line marks the overall mean dropped fraction across all IFOs and all jobs, measured to be $2.4\%$.
}
\label{fig:dropped_fraction}
\end{figure}

The GstLAL online analysis continuously processes gravitational-wave strain data from each detector in near real time.
Fig.~\ref{fig:dropped_fraction} shows the distribution of the fraction of data dropped by the analysis for each interferometer (IFO) — H1, L1, and V1 during O4.

For each strain data processing job, we track the analyzed segments and compare them against the full observing-time segments to determine the dropped fraction of data. 
The resulting fractions, aggregated across all jobs for O4, are summarized in Fig. \ref{fig:dropped_fraction} as a box-and-whisker plot. 
The box represents the interquartile range (IQR) between the 25th and 75th percentiles, the central line marks the median, and the whiskers extend to 1.5 × IQR; outliers beyond this range are shown as individual points. 

To quantify overall performance, all dropped-fraction measurements from all IFOs and all HTCondor jobs were combined and averaged.
This calculation yields a mean dropped fraction of $2.4\%$.
Interpreting the dropped fraction as the complement of analyzed observing time gives an effective analysis uptime of $97.6\%$.
Because the GstLAL online analysis is responsible for detecting signals in real time, the fraction of time during which data go unanalyzed can be viewed as a direct fractional loss in searchable volume–time.
A global dropped fraction near \qty{2}{\percent}, and per-IFO medians of the same magnitude, therefore imply that the loss in VT due to dropped data was at the few-percent level across O4.

A subset of outliers is seen for L1, traced to file corruption resulting from cooling and power-outage issues at the NEMO computing cluster where the analysis was running. 
Despite some isolated issues, all jobs analyzed more than \qty{93}{\percent} of the data, implying that the fraction of unanalyzed data remained below \qty{7}{\percent} for every IFO.
Across O4, the mean dropped fraction was 2.4\%, corresponding to 97.6\% analyzed time, and all jobs analyzed greater than 93\% of data.

The following subsection turns from overall operational stability to the production of low-latency alerts, examining how sustained uptime of the GstLAL online analysis enabled timely alert generation.

\subsection{Latency}

\begin{figure}
\includegraphics[width=\columnwidth]{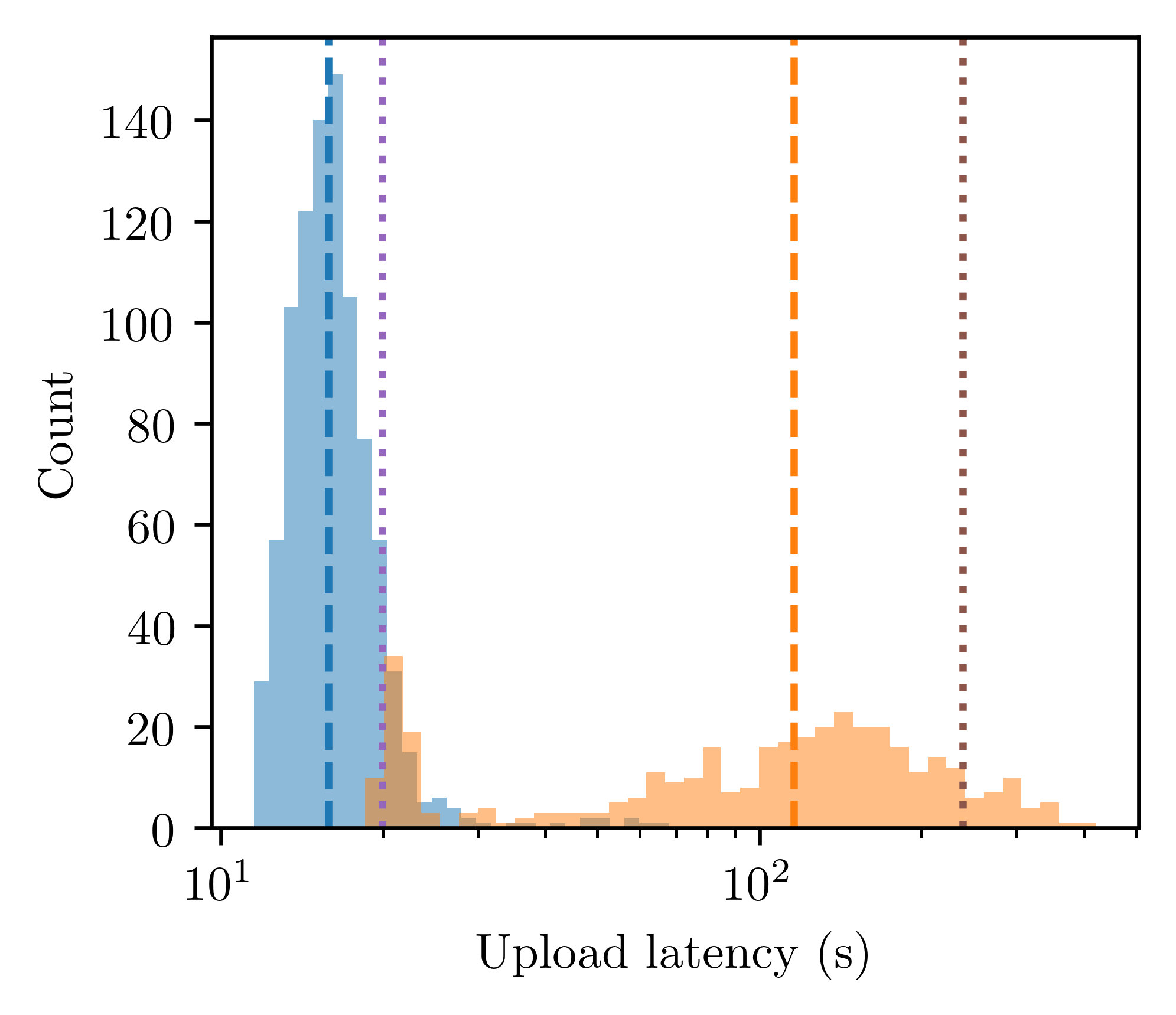}
\caption[
    Distribution of upload latencies for GstLAL online events during O4
]{%
    Distribution of upload latencies for GstLAL online events during O4. 
    The blue histogram shows the initial (non–SNR-optimized) uploads, while the orange histogram represents SNR-optimized uploads. 
    Dashed vertical lines mark, from left to right, the median and 90th-percentile latencies of the non-optimized distribution (\qty{15.8}{\second} and \qty{19.8}{\second}), followed by the median and 90th-percentile latencies of the optimized distribution (\qty{116.6}{\second} and \qty{238.3}{\second}).
    The bimodal structure of the SNR-optimized distribution arises from the use of the \textit{skymap mode} of the optimizer during part of O4, in which Virgo data were re-filtered using templates derived from Hanford–Livingston coincidences. 
    The second peak corresponds to fully optimized uploads produced by the standard SNR optimization procedure.
    \protect\footnote{Event G512720 was excluded from this figure because it was manually uploaded later.}%
}
\label{fig:o4_latency_combined}
\end{figure}

During O4 GstLAL online operations, the first candidate below a nominal FAR threshold (1 per hour), selected from aggregated triggers, is uploaded promptly.
Subsequent uploads may be delayed by a built-in geometric wait time that allows higher-SNR triggers to supersede earlier ones.
The delay is necessary to regulate the rate of uploads during periods of elevated trigger activity, preventing the low-latency infrastructure from being saturated by multiple closely spaced uploads.
As a result, the analysis prioritizes events that are more informative for follow-up.
Because sky localization accuracy improves with increasing SNR, the approach reduces unnecessary downstream processing while prioritizing uploads that yield more precise sky maps.
In addition, O4 operations used an SNR Optimizer that follows up on uploaded candidates by performing a targeted hierarchical refinement in a small neighborhood of the reported parameters \cite{Joshi:2025mdd}.
Nearby templates are dynamically generated and re-filtered to recover higher SNR and improved sky-localization inputs on timescales from seconds to minutes.

Fig.~\ref{fig:o4_latency_combined} shows the latency distributions for both the initial and SNR-optimized GstLAL uploads during O4.
The initial uploads (blue) form a narrow, sharply peaked distribution with a median latency of \qty{15.8}{\second} and a 90th percentile of \qty{19.8}{\second}, indicating that the base online pipeline consistently delivered uploads within \qty{20}{\second}.
In contrast, the SNR-optimized uploads (orange) exhibit a broader, bimodal structure with a median latency of \qty{116.6}{\second} and a 90th percentile of \qty{238.3}{\second}.
Fewer uploads appear in the SNR-optimized category because re-uploads are produced only when the optimization step yields a higher-SNR candidate.
The first peak corresponds to uploads generated while the optimizer operated in \textit{skymap mode}, in which Virgo data were rapidly re-filtered using templates from Hanford–Livingston coincidences to accelerate sky localization. 
The second peak represents fully optimized re-uploads produced by the standard optimization mode, which incurs additional computational and validation delays. 
Most optimized uploads occurred within \qty{300}{\second} of data acquisition, the deadline for finalizing the preferred event \cite{LVK_EMFollow_UserGuide_PreferredEvent} within each superevent, reflecting the scientific motivation to provide the best available sky localization for each superevent.

Overall, the latency performance summarized in Fig.~\ref{fig:o4_latency_combined} reflects a two-stage workflow in which initial uploads provide rapid alerts, followed by SNR-optimized re-uploads that refine candidate parameters on longer timescales.
The following subsection turns from low-latency alert production to event production and uploads, summarizing how these alerts contributed to superevent formation throughout O4.

\subsection{Event Production and Uploads}

The GstLAL online analysis uploaded 46569 events to GraceDB during the O4 run.
Events reported by different searches and pipelines that occur close together in time are aggregated by GWCelery \cite{gwcelery_docs} into a single superevent on GraceDB. 
Superevents that cross the significance threshold defined at the time of superevent creation, or are later determined to be astrophysically significant are marked with the label \texttt{SIGNIF\_LOCKED}.
The FAR threshold to be labelled \texttt{SIGNIF\_LOCKED} changed during the run based on the trials factor used which is dependent on the number of concurrently running pipelines. 

During O4, 271 unique superevents were labelled \texttt{SIGNIF\_LOCKED}.
Each superevent was ultimately classified by expert advocates as either:
\begin{itemize}[noitemsep, topsep=0pt]
    \item \textbf{ADVOK}: consistent with a plausible astrophysical signal, or
    \item \textbf{ADVNO}: determined to be non-astrophysical in origin (retracted).
\end{itemize}
Of the 271 \texttt{SIGNIF\_LOCKED} superevents, 254 were classified as \texttt{ADVOK} and 17 as \texttt{ADVNO}.
The GstLAL online analysis contributed uploads to 250 of the 254 \texttt{ADVOK} superevents and to 5 of the 17 \texttt{ADVNO} superevents.
A summary of GstLAL participation and event-level outcomes across the O4 \texttt{SIGNIF\_LOCKED} superevent population is provided in Table~\ref{tab:upload_stats}.

We first summarize the properties of the \texttt{ADVNO} (retracted) superevents, after which all quantitative results refer exclusively to the \texttt{ADVOK} (non-retracted) set.

\subsubsection*{ADVNO (Retracted) Superevents}
\phantomsection
\label{subsec:advno_retracted}

Most of the superevents ultimately labelled \texttt{ADVNO} did not include contributions from GstLAL, reflecting noise transients or data-quality issues that prevented the pipeline from producing viable triggers.
However, five solo-GstLAL superevents — S230708bi, S230712a, S231112ag, S240420aw, and S241126dm — were ultimately labelled \texttt{ADVNO}.
Although GstLAL was the only pipeline to upload a trigger for each of these superevents, all were retracted during advocate review due to data-quality concerns identified by detector experts or pipeline specialists.
\begin{itemize}[noitemsep, topsep=0pt]
    \item \textbf{S230708bi:} L1 exhibited data-quality problems, and the morphology of the trigger resembled a known class of instrumental artifacts.
    \item \textbf{S230712a:} Both H1 and L1 showed potential data-quality issues, no signal was recovered by online parameter estimation, and GstLAL experts judged the trigger to be inconsistent with an astrophysical signal.
    \item \textbf{S231112ag:} The event showed highly unbalanced SNR between L1 and H1 (10 vs.\ 2), along with a glitch-like feature in L1; GstLAL experts recommended retraction based on its similarity to known noise events.
    \item \textbf{S240420aw:} H1 showed glitches immediately before and after the event; detector data-quality experts advised retraction, and GstLAL experts concurred.
    \item \textbf{S241126dm:} Virgo exhibited data-quality problems at the event time, and GstLAL experts agreed with retraction.
\end{itemize}
These cases illustrate that the retractions did not stem from GstLAL behavior producing spurious triggers, but rather from underlying data-quality issues that rendered the triggers non-astrophysical.

A special case is S230920ap.
This superevent was labelled \texttt{SIGNIF\_LOCKED} solely because RAVEN \cite{piotrzkowski2022searching} reported a temporal coincidence with a Fermi trigger that was later determined to be spurious.
No significant public alert was ever issued, and the superevent's FAR remained in the non-significant regime.
For the purposes of this paper, this superevent is grouped with the \texttt{ADVNO} superevents.

\subsubsection*{ADVOK (Non-retracted) Superevents}
Of the 254 superevents labelled \texttt{ADVOK}, the GstLAL stellar-mass online analysis contributed to 250 (\(98\%\)).
The remaining four \texttt{ADVOK} superevents had no uploads from the GstLAL stellar-mass analysis for the following reasons:

\begin{itemize}[noitemsep, topsep=0pt]
    \item \textbf{S240930du:} This superevent was identified exclusively by the \textsc{cWB} burst search.  
    No matched-filter pipeline, including GstLAL found a corresponding trigger.

    \item \textbf{S250818k:} This low-significance candidate was later upgraded to \texttt{ADVOK} following independent electromagnetic observations of a new kilonova-like transient.
    The GstLAL SSM (sub-solar mass) online analysis contributed to this superevent, but the GstLAL stellar mass analysis did not produce an upload for this superevent.    

    \item \textbf{S251006dd:} A loud glitch in the Virgo detector occurred near the event time, preventing the GstLAL stellar-mass analysis from producing a reliable trigger.

    \item \textbf{S251112cm:} This superevent originated from the MBTA SSM online analysis.  
    The parameter space in which the trigger was found is not covered by the GstLAL SSM online analysis template bank, and thus no GstLAL uploads were generated.
\end{itemize}

Within the 250 \texttt{ADVOK} superevents to which the GstLAL online analysis contributed, \textbf{242} contained a GstLAL upload with a FAR below the threshold for being labelled \texttt{SIGNIF\_LOCKED}.  
Taken over the full \texttt{ADVOK} set of 254 superevents, the corresponding recovery rate for the GstLAL online analysis is \textbf{over 95\%}. 

GstLAL was also frequently the first analysis to contribute to a superevent.  
Here, ``first'' indicates that the GstLAL online analysis produced the earliest upload associated with the superevent.  
Such cases occurred for \textbf{222} of the \texttt{ADVOK} superevents to which GstLAL contributed. 
Taken over the \texttt{ADVOK} set of 254 superevents, the resulting first-contribution rate is \textbf{over 87\%}.

A related but distinct metric concerns which analysis generated the first public alert.
The first GCN Preliminary Notice corresponds to the preferred event at the time of alert issuance, as determined by the GWCelery superevent selection logic.
A preliminary notice is dispatched only after a candidate is deemed significant and all required data products are available for public dissemination.
Empirically, the latency to the first preliminary notice is approximately \qty{30}{\second}, comprising calibration and data transfer (\qty{10}{\second}), online search latency (\qty{10}{\second}), and the generation of data products, clustering, and alert distribution (\qty{10}{\second}) \cite{emfollow_alert_timeline}.
Using the identity of the pipeline associated with the first preliminary GCN as the metric, GstLAL was the first analysis for \textbf{194} superevents during O4.

Among the 250 \texttt{ADVOK} superevents to which GstLAL contributed, \textbf{137} had a GstLAL upload set as the preferred event.  
The preferred event is defined as the upload with the highest SNR among those with a FAR below the \texttt{SIGNIF\_LOCKED} threshold.  
For the \texttt{ADVOK} set, it corresponds to a preferred-event rate of \textbf{over 53\%}.  

A subset of the \texttt{ADVOK} superevents contained at least one GstLAL upload with a false alarm rate below the \texttt{SIGNIF\_LOCKED} threshold, while all uploads from other pipelines remained above threshold.  
In such cases, the classification of the superevent as \texttt{SIGNIF\_LOCKED} was supported solely by the GstLAL online analysis.  
A total of \textbf{75} superevents followed that pattern -- \textbf{over 29\%} of the \texttt{ADVOK} set.

\begin{table}[t]
\centering
\caption{
    Summary of GstLAL participation across O4 online superevents.
}
\label{tab:upload_stats}
\begin{ruledtabular}
\begin{tabular}{l c}
\textbf{Category} & \textbf{Count} \\
\hline
\multicolumn{2}{l}{\textbf{Superevent Overview}} \\
Total unique superevents & 271 \\
ADVNO (non-astrophysical) & 17 / 271 \\
ADVOK (astrophysical) & 254 / 271 \\[4pt]

\multicolumn{2}{l}{\textbf{GstLAL Coverage}} \\
ADVNO superevents with GstLAL uploads & 5 / 17 \\
ADVOK superevents with GstLAL uploads & 250 / 254 \\[4pt]

\multicolumn{2}{l}{\textbf{GstLAL Performance on ADVOK Set}} \\
Uploads below SIGNIF\_LOCKED threshold & 242 / 254 \\
Earliest upload among pipelines & 222 / 254 \\
Preferred event set by GstLAL & 137 / 254 \\
ADVOK superevents supported solely by GstLAL & 75 / 254 \\
\end{tabular}
\end{ruledtabular}
\end{table}

A consolidated overview of the statistics presented in the Event Production and Uploads subsection is given in Table~\ref{tab:upload_stats}.

\begin{figure}[htbp]
    \centering
    \includegraphics[width=\linewidth]{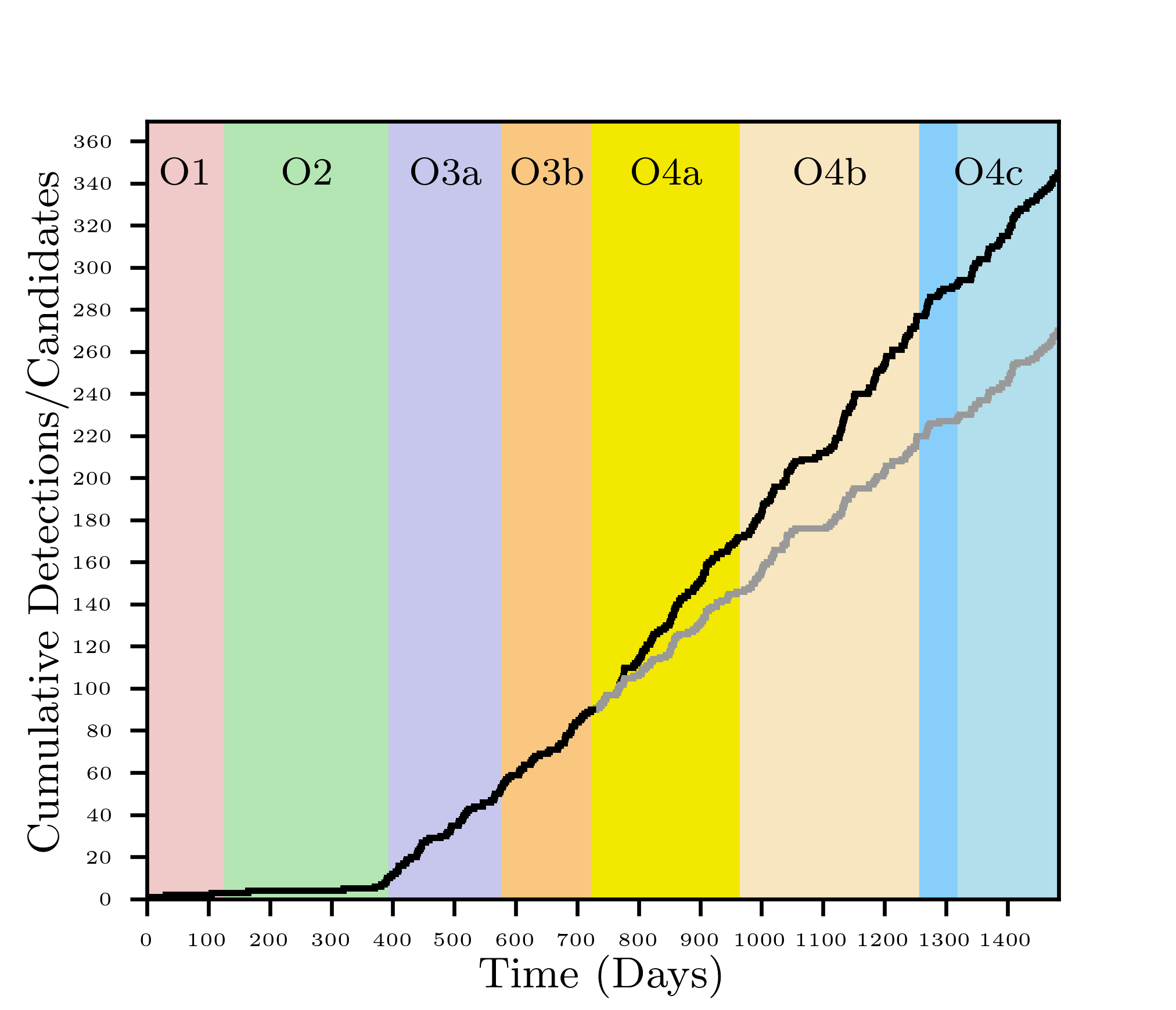}
    \caption{
        Cumulative number of detections and candidates identified by the LIGO--Virgo--KAGRA Collaboration as a function of observing time. 
        The shaded regions denote the observing runs from O1 through O4c.
        The black curve represents all cataloged events through the end of O3b, and all superevents labeled ADVOK in low-latency thereafter.         
        The grey curve shows the same distribution with all superevents removed for which the GstLAL online analysis was the sole contributor, representing the cumulative yield of the collaboration without GstLAL participation. 
    }
    \label{fig:cumulative_detections}
\end{figure}

Following the discussion of event generation and upload statistics, Fig.~\ref{fig:cumulative_detections} illustrates the cumulative number of detections identified online by the LVK as a function of observing time. 
The shaded regions mark the observing runs from O1 through O4c. 
The black curve shows all superevents labeled \texttt{ADVOK}, corresponding to events judged consistent with astrophysical signals. 
The grey curve shows the same distribution after removing superevents for which the GstLAL online analysis was the sole contributor, effectively representing the cumulative yield of the collaboration without GstLAL participation.
During O4a, 81 \texttt{ADVOK} events were identified, of which 55 were jointly recovered with other searches and 26 were unique to GstLAL. 
In O4b, the collaboration reported 105 \texttt{ADVOK} events, including 31 solo GstLAL detections. 
O4c produced 63 \texttt{ADVOK} events, 18 of which were solo.
Fig.~\ref{fig:cumulative_detections} compares the cumulative counts with and without the sole-GstLAL subset removed.

The next subsection shifts from superevent production metrics to an evaluation of the accuracy of the low-latency quantities reported by the GstLAL online analysis, comparing them against the corresponding offline estimates reported in GWTC-4.0 \cite{LIGOScientific:2025slb}.

\subsection{Accuracy of Low-Latency Quantities}

All results discussed in this subsection pertain to O4a, as only the complete O4a data are publicly available.

\subsubsection*{\pastro}

For the O4a run, 89 superevents were marked as \texttt{SIGNIF\_LOCKED}, indicating that they were promoted for expert vetting. 
Of these, 85 were identified significantly by the GstLAL online analysis, and 7 were subsequently assigned the \texttt{ADVNO} label following expert review. 
Four of these superevents — S230830b, S230808i, S230715bw, and S230622ba — had no contributions from GstLAL. 
The remaining three — S231112ag, S230712a, and S230708bi — were solo GstLAL superevents, and their retraction reasons are summarized in Sec.~\ref{subsec:advno_retracted}. 

In addition, S231123cg, which was confirmed as an astrophysical signal in GWTC-4, was classified as Terrestrial by the GstLAL online analysis. 
This misclassification occurred because the online analysis was not tuned well to recover signals in the parameter space in which this source lies.
The signal was subsequently identified with high significance in the offline analyses performed for GWTC-4.0.

Conversely, there were two instances where events identified significantly by the GstLAL online analysis were assigned lower significance offline.
The events in question were S231020bw and S230802aq, both of which were assigned lower false alarm rates offline, falling below the 2 events per day threshold required for inclusion in GWTC-4.0. 
As a result, neither event was part of the catalog despite being detected by the GstLAL online analysis.

The correspondence between GstLAL’s low-latency classifications and the final GWTC-4.0 source categories is summarized in Table~\ref{tab:confusion_matrix}.
Overall, the GstLAL online \pastro classification agrees with the final GWTC-4 source category for 92.9\% of the O4a superevents considered.

\begin{table}[htb]
    \centering
    \caption{
        Comparison between GstLAL online and GWTC-4 \pastro classifications for O4a superevents. 
        The diagonal header separates the two classification sources.
        The Terrestrial--Terrestrial entry is intentionally omitted, as reporting agreement for events classified as non-astrophysical carries no scientific meaning in this context.
    }
    \begin{tabular}{lccc}
        \hline\hline
        \diagbox{\textbf{GWTC-4}}{\textbf{GstLAL Online}} & \textbf{NSBH} & \textbf{BBH} & \textbf{Terrestrial} \\
        \hline
        NSBH          & 1 & 0  & 0 \\
        BBH          & 0 & 78 & 1 \\
        Terrestrial  & 0 & 5  & -- \\
        \hline
    \end{tabular}
    \label{tab:confusion_matrix}
\end{table}



\subsubsection*{FAR Comparison}

\begin{figure}[htbp]
    \centering
    \includegraphics[width=\linewidth]{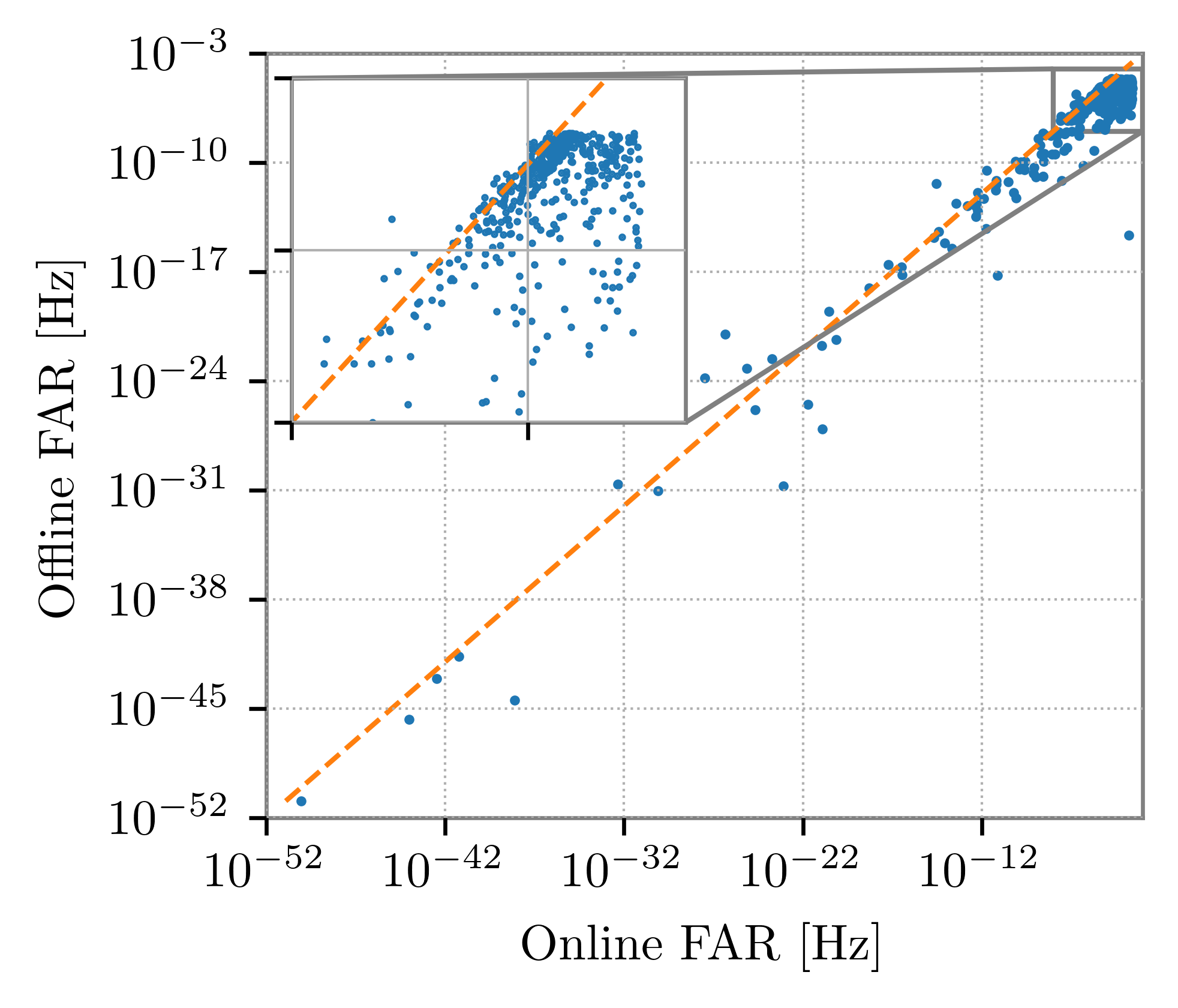}
    \caption{
        Comparison of the online and offline FARs for all GWTC-4.0 superevents GstLAL participated in O4a. 
        Each point represents a single superevent, with the dashed line indicating equality between the two estimates. 
        Most superevents cluster near the one–to–one line, demonstrating broad agreement between the online GstLAL pipeline and the offline processing.
        Offline reprocessing incorporates updated calibration, extended background, and full data–quality vetoes, leading to systematically more accurate FAR estimates.
    }
    \label{fig:far_comparison}
\end{figure}

A comparison of online and offline FARs for all GWTC-4.0 superevents in Fig.~\ref{fig:far_comparison} shows that the majority of superevents to which GstLAL contributed lie close to the one-to-one line.
Rather than focusing on numerical agreement in FAR values, a more operationally relevant question is whether offline reprocessing identifies astrophysically significant events that were not recovered at comparable significance online.
Using a FAR threshold of $1\,\mathrm{yr}^{-1}$ to define significance, the catalog contains $86$ astrophysically plausible candidates, all of which had corresponding low-latency GstLAL uploads.
Of these, $76$ were given the ADVOK label, while the remaining $10$ crossed the threshold only after offline reprocessing.
One candidate in this set was ultimately classified as being of instrumental origin, despite being recovered by the GstLAL online analysis with a FAR below $1\,\mathrm{yr}^{-1}$.
In all such cases, the online GstLAL analysis produced low-latency candidates with higher FARs that were subsequently promoted to significance by offline analyses benefiting from extended background accumulation, updated calibration, and refined data-quality information.

\subsection{Missed Events}

Although nine superevents ultimately included in the catalog did not get labelled as \texttt{SIGNIF\_LOCKED} online, the GstLAL online analysis nevertheless produced corresponding low-latency uploads for each of them. 
In all such cases, the online GstLAL uploads were recovered with lower significance than the threshold required for a \texttt{SIGNIF\_LOCKED} designation. 
These events subsequently achieved higher significance in the offline analysis, where the analyses benefit from improved calibration, refined data-quality information, and access to the full data stretch, so the upward shift in significance is consistent with expectations for offline reprocessing.

\subsection{Live Injections}

\begin{table}[htb]
    \centering
    \caption{
        Confusion matrix comparing injection source classes to GstLAL online \pastro classifications for live injections.
        Rows correspond to the true source class determined from injection metadata, while columns indicate the source class with the maximum \pastro component assigned by the online analysis.
        Only VTInjection candidates with FARs below \qty{1} per \qty{6}{months} and occurring in time segments passing final data-quality requirements are included.
    }
    \begin{tabular}{lcccc}
        \hline\hline
         & \multicolumn{4}{c}{\textbf{GstLAL Online Classification}} \\
        \cline{2-5}
        \textbf{True Source Class} & BNS & NSBH & BBH & Terrestrial \\
        \hline
        BNS          & 1037 & 2130 & 1     & 37  \\
        NSBH         & 103  & 2141 & 318   & 15  \\
        BBH          & 0    & 782  & 67914 & 0   \\
        \hline
    \end{tabular}
    \label{tab:vtinjection_confusion_matrix}
\end{table}

Live software injections were enabled and streamed with the detector data to allow real-time monitoring of injection recovery by the online searches.
The GstLAL online analysis processed these injections and uploaded associated candidates to GraceDB Playground under the \texttt{VTInjection} search tag.

In this paper, we consider VTInjection uploads that would have satisfied the \texttt{SIGNIF\_LOCKED} criteria applicable during O4b if they corresponded to astrophysical signals.
Specifically, we select VTInjection candidates with FARs below \qty{6.34e-8}{\hertz}.
For a given injection, multiple candidates may be uploaded by the analysis.
Candidates within \qty{1}{\second} of each other are clustered, and the candidate with the highest SNR is selected as the preferred event, consistent with the procedure used to identify preferred events within a superevent.

For each preferred candidate, the \pastro values reported by the online analysis are used to determine the assigned source class, defined as the class corresponding to the maximum \pastro component.
The corresponding injection is identified using the injected time as a proxy, requiring the recovered candidate time to be within \qty{1}{\second} of the injection time.
Only candidates occurring in time segments that pass final data-quality requirements are included.

The true source class of each injection is determined from the injected component masses using standard classification criteria.
Table~\ref{tab:vtinjection_confusion_matrix} presents a confusion matrix comparing the injection truth classes with the source classes assigned by the GstLAL online \pastro classification.
Overall, \qty{95.3}{\percent} of the injections considered are assigned the correct source class.
This metric reflects the real-time source classification behavior of the GstLAL online analysis.

\section{Operational Experience and Automation}
\label{sec:experience}

\subsection{Operational Model}

The GstLAL online analysis operates within a rota-based human oversight framework that provides continuous monitoring during observing runs.
Each weekly rota-shift assigns three responders with tiered roles -- a primary lead (Responder 1), an intermediate responder (Responder 2), and a trainee (Responder 3).
The tiered structure provides operational stability by ensuring that each shift includes both experienced operators and members gaining expertise, facilitating knowledge transfer and strengthening long-term resilience across observing runs, while also incorporating fresh perspectives that can identify improvements and refine procedures.

Responders monitor the analysis through dashboards, HTCondor job status panels, detector state, and automated alert channels.
Core responsibilities include monitoring DAG execution, identifying stalled or held jobs, verifying data ingestion stability, and ensuring that uploads to GraceDB occur as expected.
Responders also perform routine DAG maintenance, including restarting jobs, clearing failures, and validating propagation of configuration or container updates.
During interventions, responders set the appropriate downtime in monitoring systems to prevent spurious alerts and to communicate information about said intervention.
Responders participate in coordination meetings, provide status updates, and contribute to weekly rota reports that summarize shift activity and maintain continuity across shifts.

Responder roles are differentiated by responsibility.
Responder~1 leads operations, manages alert response, coordinates debugging, oversees follow-up for \texttt{SIGNIF\_LOCKED} events, and provides updates during meetings.
Responder~2 assists failure diagnosis and monitoring by tracking HTCondor logs, identifying anomalies, and supporting superevent follow-up.
Responder~3 focuses on training by shadowing the senior responders, monitoring dashboards, subscribing to alerts and contributing to documentation.

Human oversight remains central to the reliability of the online analysis.
Responders routinely inspect the DAG structure, monitor injection recovery dashboards, track resource usage across nodes, and assess overall analysis health.
\texttt{SIGNIF\_LOCKED} superevents trigger additional checks such as Q-scan inspection, verification of detector behavior around the event time, and confirmation that background estimates and uploads are consistent with expectations.
The outlined procedures provide real-time validation of automated results by trained analysts.

Extended observing periods introduce operational challenges, including responder fatigue, particularly during night-time alerts or prolonged debugging efforts.
Debugging under time pressure is difficult, as diagnosing failures often requires coordinating across multiple systems including Condor logs, node stability, network connectivity, and data distribution services.
The pipeline's DAG structure adds further fragility since failures in upstream jobs can cascade into larger workflow interruptions.
Documentation maintenance can lag during busy shifts, as urgent tasks take priority, while continuous monitoring and alert streams contribute to operational fatigue even during quiet periods.

The rota model provides structure, redundancy, and continuity for the GstLAL online analysis, ensuring that human oversight remains integral to the stability and reliability of real-time gravitational-wave detection.

The operational model outlined above highlights the extent of human oversight required to maintain the stability of the online analysis during observing runs.
The layered responder structure, continuous monitoring, and frequent intervention ensure reliability but also place a substantial burden on operators, especially during periods of high activity or extended shifts.
The demands motivated a concerted effort during O4 to reduce routine monitoring tasks, streamline responder workload, and alleviate operational fatigue through targeted automation.
The next subsection describes the automation tools and procedures developed, and discusses how they reduced operational overhead while preserving the reliability of the online analysis.

\subsection{Automation Efforts}

During O4, \textit{Sherlog} was developed as a targeted automation system for the GstLAL online analysis.
It handles well-defined, repetitive tasks within the rota-based operational model while preserving human oversight.
Its scope is limited to high-value operations that historically required frequent manual checks, enabling a more consistent and efficient workflow.

A primary capability of \textit{Sherlog} is a daily automated error scan designed to identify silent failures not captured by standard monitoring systems.
Silent failures can degrade analysis stability by allowing jobs to stall or error states to persist without generating alerts.
\textit{Sherlog} performs a systematic scan of analysis logs, aggregates detected issues, and posts structured summaries -- including job identifiers, error types, and timestamps, to a responder Mattermost channel \cite{mattermost}.
This process provides a consistent overview of analysis health and enables early identification of recurring issues without manual log inspection.

\textit{Sherlog} also performs real-time diagnostics of failures as they occur.
It monitors for stalled jobs, repeated errors, and node-level issues, including cases where compute nodes become unavailable.
The diagnostics provide responders with immediate, actionable information about the state of the analysis, reducing the need for direct interaction with cluster infrastructure during time-critical interventions.
Once a failure mode has been identified, \textit{Sherlog} can be extended to recognize the condition automatically, reducing the need for repeated manual diagnosis.

In addition to monitoring and diagnostics, \textit{Sherlog} automates several routine operational procedures.
It manages the weekly takedown of the online analysis prior to scheduled detector maintenance by monitoring detector state and initiating shutdown when required, while reporting status updates throughout the process.
It also automates the setup and execution of the weekly re-whitening workflow, launching the process at empirically determined times and handling the required preparation steps.
Such automations ensure consistent execution of critical maintenance operations and remove recurring manual tasks.

Targeted automation was also applied to specialized procedures that directly affect analysis integrity.
An automated count-removal tool was developed to eliminate signal contamination from background estimates following the identification of significant events.
The tool performs the full sequence of operations, including identifying the appropriate analysis build, executing the removal, verifying the result, and reporting progress.
Standardizing the count-removal procedure improves consistency and reduces the risk of error during periods of high activity or responder fatigue.

Additional automation supports resilience through rolling backups of key analysis products.
The system maintains up-to-date copies of critical files and performs periodic cleanup to manage storage usage.
These backups enable rapid recovery from failures such as node crashes or data corruption, reducing downtime and minimizing the need for manual reconstruction of analysis state.
Although implemented independently of \textit{Sherlog}, this functionality contributes directly to the stability and robustness of the online workflow.

Together, the automation efforts form a layer that enhances the reliability of the GstLAL online analysis while reducing the frequency of routine manual intervention.
\textit{Sherlog} handles repetitive monitoring, standardizes operational procedures, and preserves knowledge of known failure modes within the system.

Automation does not replace the need for human oversight.
Many operational decisions require expert judgment, contextual awareness, and coordination across systems.
The effectiveness of the online analysis therefore continues to depend on the experience and attentiveness of responders, particularly in diagnosing novel issues and managing complex situations.

The following subsection examines the human aspects of online operations and discusses how responder practices, coordination, and experience continue to shape the overall performance and reliability of the analysis.

\subsection{Human Aspects}

Even with extensive automation in place, the stability and responsiveness of the online analysis continued to depend critically on the human responders who monitored, interpreted, and acted on system behavior throughout O4.
Responders provided contextual awareness, operational experience, and scientific judgment that could not be replicated by automated checks, particularly in situations where multiple subsystems behaved unexpectedly or requiring rapid coordination across teams.
The interaction between responders and automation shaped the overall effectiveness of the analysis, with tools such as \textit{Sherlog} reducing routine workload while leaving higher-level decision-making to responders.

During O4, the handling of development tasks revealed the importance of clear delegation within the online analysis team.
When a problem was identified, assigning a specific individual to investigate and implement a solution led to more efficient resolution than distributing responsibility across multiple responders.
Reducing that individual's operational duties during the development period enabled sustained focus on diagnosing root causes and implementing stable fixes.
Newer members were able to gain experience through direct engagement with the analysis infrastructure, in some cases leading to additional improvements and feature development.

Documentation emerged as another critical area where operational experience revealed gaps.
Although automation captured many recurring failure modes, responders often relied on accumulated experience and informal knowledge when diagnosing unfamiliar issues or interpreting ambiguous system behavior.
The uneven distribution of institutional knowledge led to variability in response strategies and time-to-diagnosis across shifts.

O4 also highlighted challenges in prioritization and response strategies.
While automated systems surfaced actionable failures, responders were still required to assess their severity and determine appropriate actions.
In the absence of consistently applied prioritization schemes, differences in interpretation occasionally led to delays in response or variation in handling similar issues.

Responder fatigue and workload distribution were recurring concerns throughout O4.
The irregular shift schedule, nighttime responsibilities, and sustained attention required to monitor a complex analysis placed cumulative strain on responders over extended periods.
Even with automation addressing many routine tasks, human oversight remained essential, and the associated workload contributed to operational fatigue.

The operational demand was further increased by the need to manage multiple analyses running concurrently in different configurations across O4.
Responders were required to track the state, performance, and data products of several analyses simultaneously, with configurations evolving between different phases of the observing run.
The complexity introduced substantial cognitive overhead and compounded the challenges associated with sustained monitoring and rapid response.

The experience gained during O4 highlighted both the strengths of the online analysis framework and the practical limitations encountered during long-duration operation.
The interaction between automation, human oversight, and targeted development provided a stable foundation, but also revealed organizational and procedural challenges in how issues were identified, prioritized, and resolved.
These observations motivate the improvements and priorities discussed in the following section.
\section{Lessons for the Next Observing Run}
\label{sec:lessons}
\subsection{Pipeline improvements}

Supporting the higher event rates and broader science goals anticipated for next observing runs requires targeted improvements to the GstLAL online analysis.
The limitations identified during O4 highlight several development priorities, including reducing alert latency, improving reliability, and strengthening scalability across heterogeneous computing environments.
Improvements to monitoring and diagnostic infrastructure are also required to reduce responder burden and enable faster, more consistent interpretation of system behavior.

Scalability remains a central challenge for the next observing runs.
The diversity of analyses operated during O4 highlighted limitations in how configurations were managed, deployed, and synchronized across clusters.
Supporting multiple concurrent analyses, each with distinct parameter spaces and runtime requirements, demands a more scalable and reproducible framework.
Enhancements to configuration management and automated version tracking will reduce operational complexity and ensure consistent behavior across deployments.
Improved tooling for rapid and low-risk instantiation of new analyses will also facilitate commissioning and testing during live observing periods.

Strengthening monitoring and diagnostic tools is critical for maintaining stable operations at higher observing cadence.
While \textit{Sherlog} provided effective first-order triage during O4, more granular and automated diagnostic capabilities are required to interpret complex and evolving failure modes.
Developing clearer visualizations, automatic cross-checks, and improved root-cause indicators will enable faster and more consistent identification of issues.
Automated recovery mechanisms and predictive diagnostics will reduce the frequency and impact of interruptions.
Improvements to \textit{Sherlog} will also reduce reliance on accumulated operational experience and make the online analysis more accessible to a broader range of contributors.

Collectively, the enhancements reflect the evolution of the GstLAL online analysis toward the scalability, reliability, and operational stability required for the next observing runs.

\subsection{Automation roadmap}

The automation framework developed during O4 provides a foundation for further expansion in future observing runs.
Building on the pipeline improvements described in the previous subsection, the next stage of work focuses on strengthening automated monitoring, expanding intelligent diagnostics, and reducing manual intervention during critical periods of observing.

A primary direction involves expanding \textit{Sherlog's} diagnostic capabilities beyond node-level failures.
One enhancement includes identifying switch-level failures, where loss of a network switch can disrupt multiple nodes simultaneously, enabling more accurate triage without manual pattern recognition.
Extending diagnostics to inter-node communication issues will also improve detection of intermittent job stalls and degraded throughput.
The roadmap also includes expanding the library of encoded failure modes based on issues encountered during extended observing periods, allowing \textit{Sherlog} to recognize and respond to a broader range of recurring conditions automatically.
In addition, enabling \textit{Sherlog} to automatically relaunch analyses following maintenance, provided required inputs such as SVD filters are available, will reduce downtime at the boundaries of scheduled maintenance windows and improve continuity of operations.

A longer-term direction involves integrating an automated assistant capable of interpreting system behavior beyond pre-defined failure modes.
Unlike \textit{Sherlog}, which relies on explicitly encoded rules to identify and respond to known conditions, such an assistant would support more flexible reasoning by synthesizing information across logs, system states, and recent operational history.
The assistant would enable responders to query the system in natural language, receive contextualized diagnoses, and translate high-level intent into coordinated operational actions.
Over time, such an assistant would be expected to manage an increasing fraction of routine online operations, reducing real-time cognitive load on responders and allowing greater focus on development, methodological improvements, and detailed analysis of pipeline performance.
Such an assistant would ultimately shift responder effort from reactive maintenance toward proactive innovation, improving both operational reliability and long-term scientific productivity.

Together, the efforts point toward a more autonomous and resilient GstLAL online analysis.
Increasing the ability of the pipeline to monitor, diagnose, and recover from routine issues will reduce operational burden while maintaining stable performance under higher observing cadence.
\section{Conclusions}

The GstLAL online analysis performed reliably throughout O4, maintaining high uptime, recovering most astrophysical candidates, and delivering low-latency results for real-time alerts.
The quantitative results confirm that the pipeline met its design objectives, delivering rapid alerts, providing reliable classifications, and consistently supporting superevent identification.
The operational lessons from O4 point toward expanded automation, improved stability across multiple configurations, and shortened paths to catalog-ready analysis results in the next observing run.
The developments will further enhance the readiness of the GstLAL online analysis for higher event rates as we delve deeper into the field of GW astronomy.

The results presented in Sec.~\ref{sec:performance} provide a comprehensive assessment of the GstLAL online analysis during O4. 
The latency measurements show that initial uploads were consistently delivered within tens of seconds and that SNR-optimized re-uploads were produced within a few minutes, meeting the requirements for rapid alert generation and sky-localization refinement. 
The uptime and dropped-fraction statistics demonstrate that the analysis operated with an effective duty cycle above \qty{93}{\percent}, ensuring continuous low-latency coverage across all observing periods. 
The event-production metrics highlight the role of the GstLAL online analysis in O4, with the online analysis contributing to over \qty{98}{\percent} of all \texttt{ADVOK} superevents, providing the first upload in most cases, and serving as the sole low-latency support for nearly a third of the \texttt{ADVOK} set. 
Comparisons between online and offline quantities—including FAR, \pastro, and source classification show broad consistency between the two analyses.
Finally, superevents that did not achieve \texttt{SIGNIF\_LOCKED} status online but were recovered with higher significance offline reflect the expected differences rather than shortcomings of the GstLAL online analysis. 
Together, the results demonstrate that the GstLAL online analysis delivered rapid, robust, and scientifically reliable performance throughout O4. 

\begin{acknowledgments}
This work has made use of data, software and/or web tools obtained from the Gravitational Wave Open Science Center (\texttt{https://www.gw-openscience.org/}), a service of LIGO Laboratory, the LSC and the Virgo Collaboration.
We especially made heavy use of the LVK Algorithm Library.
LIGO was constructed by the California Institute of Technology and the Massachusetts Institute of Technology with funding from the United States National Science Foundation (NSF) and operates under cooperative agreements PHYS-$0757058$ and PHY-$0823459$.
In addition, the Science and Technology Facilities Council (STFC) of the United Kingdom, the Max-Planck-Society (MPS), and the State of Niedersachsen/Germany supported the construction of aLIGO and construction and operation of the GEO600 detector.
Additional support for aLIGO was provided by the Australian Research Council.
Virgo is funded, through the European Gravitational Observatory (EGO), by the French Centre National de Recherche Scientifique (CNRS), the Italian Istituto Nazionale di Fisica Nucleare (INFN) and the Dutch Nikhef, with contributions by institutions from Belgium, Germany, Greece, Hungary, Ireland, Japan, Monaco, Poland, Portugal, Spain.

This material is based upon work supported by NSF's LIGO Laboratory which is a major facility fully funded by the National Science Foundation. 
The authors are grateful for computational resources provided by the LIGO Lab cluster at the LIGO Laboratory and supported by PHY-$0757058$ and PHY-$0823459$, the Pennsylvania State University's Institute for Computational and Data Sciences gravitational-wave cluster (RRID:SCR\_025154), and the University of Wisconsin Milwaukee Nemo and supported by PHY-$1626190$ and PHY-$2110594$.
This work was supported by PHY-$2513358$, PHY-$2308881$, OAC-$2346596$, OAC-$2201445$, OAC-$2103662$.

\par\AckText
\end{acknowledgments}

\bibliography{references}

\end{document}